\def\year{2022}\relax
\documentclass[letterpaper]{article} 
\usepackage{aaai22}  
\usepackage{times}  
\usepackage{helvet}  
\usepackage{courier}  
\usepackage[hyphens]{url}  
\usepackage{graphicx} 
\usepackage{natbib}  
\usepackage{caption} 
\DeclareCaptionStyle{ruled}{labelfont=normalfont,labelsep=colon,strut=off} 
\usepackage{algorithm}
\usepackage{algorithmic} 
\usepackage{subcaption}
\usepackage{graphicx}
\usepackage{float}
\usepackage[dvipsnames]{xcolor}
\usepackage{url} 
\usepackage{amsmath}

\usepackage{newfloat}
\usepackage{listings}
\floatstyle{ruled}
\newfloat{listing}{tb}{lst}{}
\floatname{listing}{Listing}
\title{Understanding human mobility patterns in Chicago: an analysis of taxi data using clustering techniques\thanks{Project done as part of the course on Mathematical Tools for Data Science at NYU Center for Data Science.}}
\author {
     Harish Chauhan\thanks{All authors contributed equally to this work.},
     Nikunj Gupta\footnotemark[2],
     Zoe Haskell-Craig\footnotemark[2]
 }
\affiliations {
    New York University\\
    \texttt{\{hpc4473, nikunj.gupta, zjh235\}@nyu.edu} 
}

\begin{document}

\maketitle

\section{Motivation}
Understanding human mobility patterns is important in applications as diverse as urban planning, public health and political organizing. One rich source of data on human mobility is taxi ride data. Using the city of Chicago as a case study, we examine data from taxi rides in 2016 with the goal of understanding how neighborhoods are interconnected. This analysis will provide a sense of which neighborhoods individuals are using taxis to travel between, suggesting regions to focus new public transit development efforts. Additionally, this analysis will map traffic circulation patterns and provide an understanding of where in the city people are travelling from and where they are heading to - perhaps informing traffic or road pollution mitigation efforts. For the first application, representing the data as an undirected graph will suffice. Transit lines run in both directions so simply a knowledge of which neighborhoods have high rates of taxi travel between them provides an argument for placing public transit along those routes. However, in order to understand the flow of people throughout a city, we must make a distinction between neighborhood from which people are departing and the areas to which they are arriving - this requires methods that can deal with directed graphs. All developed codes can be found at {\color{magenta}\url{https://github.com/Nikunj-Gupta/Spectral-Clustering-Directed-Graphs}}. 

\section{Data}
Data was taken from the taxi trips for 2016, reported to the City of Chicago, and available on Kaggle. (\url{https://www.kaggle.com/datasets/chicago/chicago-taxi-rides-2016?select=data\_dictionary.csv}) This data contained information on pick-up community area and drop-off community area, the community areas where taxi trips began and ended. A total of almost 1 300 000 observations were included. We converted this into a graph considering the community areas as nodes and mean travel time between nodes as the weighted edges. As such, we established a directed graph consisting of 77 nodes representing the taxi traffic between areas of Chicago.

\section{Methodology}

\subsection{Clustering methods for undirected graphs}
Several algorithms exist to cluster nodes of an undirected graph. We modified our directed graph to an undirected graph by considering any trip between two community areas to contribute to the edge weight, independent of the direction of the trip. We then applied the following algorithms to establish cluster in the graph:

\paragraph{Spectral clustering:} 

Spectral graph theory associates spectral properties of matrices to undirected graphs. We define the graph Laplacian L, given by L = I - $D^{-1}$W, where D is the diagonal matrix of degrees and W is the weighted adjacency matrix. Matrix L is real and symmetric, hence diagonalizable, and has real eigenvalues and eigenvectors,  with zero as an eigenvalue (always) and the vector of all ones as the corresponding eigenvector. The key incentive of spectral clustering is to group the nodes of a graph into two or more clusters based on relating them to the spectrum of the graph Laplacian. The spectral clustering algorithm involves the computation of the k smallest eigenvalues and the use of all k corresponding eigenvectors to cluster graph nodes into k clusters.  

\begin{algorithm}[H]
\hfill 
\begin{enumerate}
    \item \textbf{Input:} Adjacency matrix $W \in R^{n \times n}_+$ of a directed graph; 
    \item \textbf{Parameters:} $ k\in \{2,3,..n\}$
    \item Compute the graph Laplacian L;
    \item Find the k first eigenvectors and store them as the columns of a matrix
    \item Consider each row of the above matrix as a point in $R^k$ and cluster these points using a k-means algorithm 
    \item \textbf{Output:} estimation of cluster membership; 
\end{enumerate}
\caption{Spectral clustering algorithm \label{alg:sc}}
\end{algorithm}

\paragraph{Leiden Algorithm:}
The Leiden algorithm uses the Constant Potts Model defined as $H = \sum [e_c - \gamma {\binom{n_c}{1}}]$, where $n_c$ is the number of nodes in cluster $c$, $e_c$ is the number of edges in $c$, and $\gamma$ is a tuning parameter. The Leiden algorithm works by iterating through the following steps \cite{r:Leiden}:

\begin{algorithm}[H]
\hfill 
\begin{enumerate}
    \item Start with a singleton partition.
    \item Moves nodes from one cluster to another in order to find a partition.
    \item Refine the clusters, which may result in splitting the clusters into multiple subcluster by merging well-connected nodes. 
    \item Aggregate clusters to form a single node in the subsequent iteration of the algorithm (now on a smaller graph).
\end{enumerate}
\caption{Leiden Algorithm \label{alg:leiden}}
\end{algorithm}

The Leiden algorithm is implemented through the igraph package available for python. 

\paragraph{Walktrap:}
The walktrap algorithm developed in the igraph package uses an algorithm proposed in \cite{pons2005computing}. This approach uses a partition of the graph to define clusters.
The intuition behind this algorithm is that a random walk tends to get "trapped" in the densely connected subregions of the graph corresponding to a cluster. As such, Pons and Latapy create a notion of distance, $r$, between nodes i and j in terms of the probability of going from i to j through a random walk of length t. The walktrap algorithm then works as follows:

\begin{algorithm}[H]
\hfill 
\begin{enumerate}
    \item Choose two clusters based on the distance $r$ between them. 
    \item Merge these two clusters to form a single cluster, and create a new partition of the graph. In this algorithm, only adjacent clusters are merged. 
    \item Update the distances between clusters. 
\end{enumerate}
\caption{Walktrap Algorithm \label{alg:walktrap}}
\end{algorithm}

\subsection{Clustering methods for directed graphs}

Directed graphs have their vertices connected by directional edges. Unlike undirected graphs, the direction of the edges between two vertices convey distinct information, in this case the direction of taxi travel. Hence the weight matrix representing the edge is not symmetric. In our case, the value in $W_{ij}$ represents the time taken to go from neighborhood $i$ to $j$. Given any directed graph, we can symmetrize the graph to an undirected graph by dropping the direction of each edge. We can then use any of the numerous partitioning methods for undirected graphs. The problem with this approach is that it loses key information.

There are also a wide variety of clustering methods for directed graphs. Generally speaking, algorithms can be partitioned into two categories: given a directed graph, either we build a related undirected graph and apply known clustering algorithms to it or we work directly on the directed graph and formulate new methods from scratch. In this section we explain some of the methods we implemented on the directed graph we obtained for the data on Chicago taxi trips. 

\paragraph{Simple Transformation:} In this case, a new undirected graph with symmetric adjacency matrix is created from the original directed graph using a simple matrix transformation (shown below). Then clustering methods for undirected graphs are applied to this transformed weight matrix.     

\[W_u = W + W^T\]

This transformation simply ignores the directionality of edges. Hence, it should be avoided if the goal is to take the directionality of edges into account. 

\paragraph{Bibliometric Symmetrization:} Satuluri et al. proposed this type of transformation from non-symmetric to symmetric weight matrix \cite{satuluri2011symmetrizations}. The transformation is defined as follows: 

\[W_u = W^T W + W W^T\]
    
The reason that it is able to find patterns is that $W^T W$ measures common in-neighbours and $W W^T$ measures common out-neighbours for every pair of nodes in the directed graph. As there does not seem to be any obvious reason for leaving out either in-neighbours or out-neighbours, a sum of both matrices is taken so as to account for both. 

\paragraph{Spectral clustering using Chung’s directed Laplacian:} A straightforward generalization of the Laplacian matrix to the directed case can lead to a non-symmetric matrix because the adjacency matrix of a directed graph is not symmetric. This makes spectral analysis more difficult (complex eigenvalues, defective matrices, etc). Fan Chung proposed a symmetrized version of the Laplacian that applies to directed graphs \cite{chung2005laplacians}. This method was further developed by David Gleich in \cite{gleich2006hierarchical} to cluster nodes in directed graphs. Given a strongly connected directed, aperiodic graph G = (V, E) and a weight matrix $W \in R^{n \times n}_+$, Chung's Directed Laplacian is given by 

\[L(G) = I - \frac{1}{2} ( \pi^{1/2} P \pi^{-1/2} +  \pi^{-1/2} P^T \pi^{1/2})\]

where, P the transition matrix of the graph given by $P = D^{-1}W$ and D is the out degree matrix $D = W \mathbf{1}$ (\textbf{1} is the column vector of 1’s).

\begin{algorithm}[H]
\hfill 
\begin{enumerate}
    \item \textbf{Input:} Adjacency matrix $W \in R^{n \times n}_+$ of a directed graph; 
    \item \textbf{Parameters:} $ k\in \{2,3,..n\}$
    \item Compute the graph's Chung’s directed Laplacian L;
    \item Find the k first eigenvectors and store them as the columns of a matrix
    \item Consider each row of the above matrix as a point in $R^k$ and cluster these points using a k-means algorithm 
    \item \textbf{Output:} estimation of cluster membership; 
\end{enumerate}
\caption{Chung’s directed Laplacian (CDL) Spectral clustering \label{alg:chung}}
\end{algorithm}

\paragraph{SVD spectral clustering:} Sussman et al propose an algorithm based on the singular value decomposition of the adjacency matrix \cite{sussman2012consistent}. The algorithm includes the latent feature dimension (d), which should be number of singular values of W significantly greater than 0. The adjacency matrix W is expected to be well approximated by a matrix of rank d. The main difference with classical spectral clustering algorithms is that this algorithm uses the non-symmetric adjacency matrix instead of a Laplacian matrix. An advantage of SVD spectral clustering is that it is very general in the type of clusters it is able to detect. 

\begin{algorithm}[H]
\hfill 
\begin{enumerate}
    \item \textbf{Input:} Adjacency matrix $W \in R^{n \times n}_+$ of a directed graph; 
    \item \textbf{Parameters:} $ d\in \{1,2,3,..n\}$, $ k\in \{2,3,..n\}$
    \item Compute the singular value decomposition W = US$V^T$ where S has decreasing main diagonal;
    \item Let $\tilde{U}$ and $\tilde{V}$ be the first d columns of U and V , let $\tilde{S}$ contain the first d columns and first d rows of S;
    \item Define $\tilde{Z}$ = [$\tilde{U}$ $\tilde{S}^{1/2}$ \textbar $\tilde{V}$ $\tilde{S}^{1/2}$] by concatenation; 
    \item Cluster the rows of $\tilde{Z}$ using a k-means algorithm with Euclidean distance. 
    \item \textbf{Output:} estimation of cluster membership; 
\end{enumerate}
\caption{SVD spectral clustering \label{alg:svd}}
\end{algorithm}

\paragraph{Spectral clustering algorithm based on random walk:} This algorithm mainly extract information from eigenvectors and eigenvalues of the transition matrix of a directed graph. The method focuses on finding the second largest eigenvalue of transition matrix P in terms of magnitude and then look at the components of the corresponding eigenvector to partition the network into two clusters. Finding this second eigenvector seems to be sufficient to partition the network in more than two clusters.

\begin{algorithm}[H]
\hfill 
\begin{enumerate}
    \item \textbf{Input:} $W \in \{0, 1\}^{n \times n}$;
    \item \textbf{Parameters:} $d \in \{1,2,...,n\}, k \in \{2,3,...,n \} $
    \item Compute the transition matrix P = $D^{-1}$W
    \item Calculate the second eigenvector u of P and calculate vector v such that $v_i =Re(u_i)+Im(u_i),$ for $1 \leq i \leq n;$
    \item Apply the Gaussian kernel function to the components of v and store the result in a vector w;\\
    \[w_i = \Phi_v(v_i)\]
    \item Cluster the components of w into k clusters (for instance by using k-means algorithm). 
    
    \item \textbf{Output:} estimation of cluster membership;
\end{enumerate}
\caption{Random Walk Spectral Clustering Algorithm \label{alg:randwalk}}
\end{algorithm}

\section{Results and Conclusion} 

\begin{figure}[h]
\centering
\begin{subfigure}[b]{0.15\textwidth}
    \includegraphics[width=\textwidth]{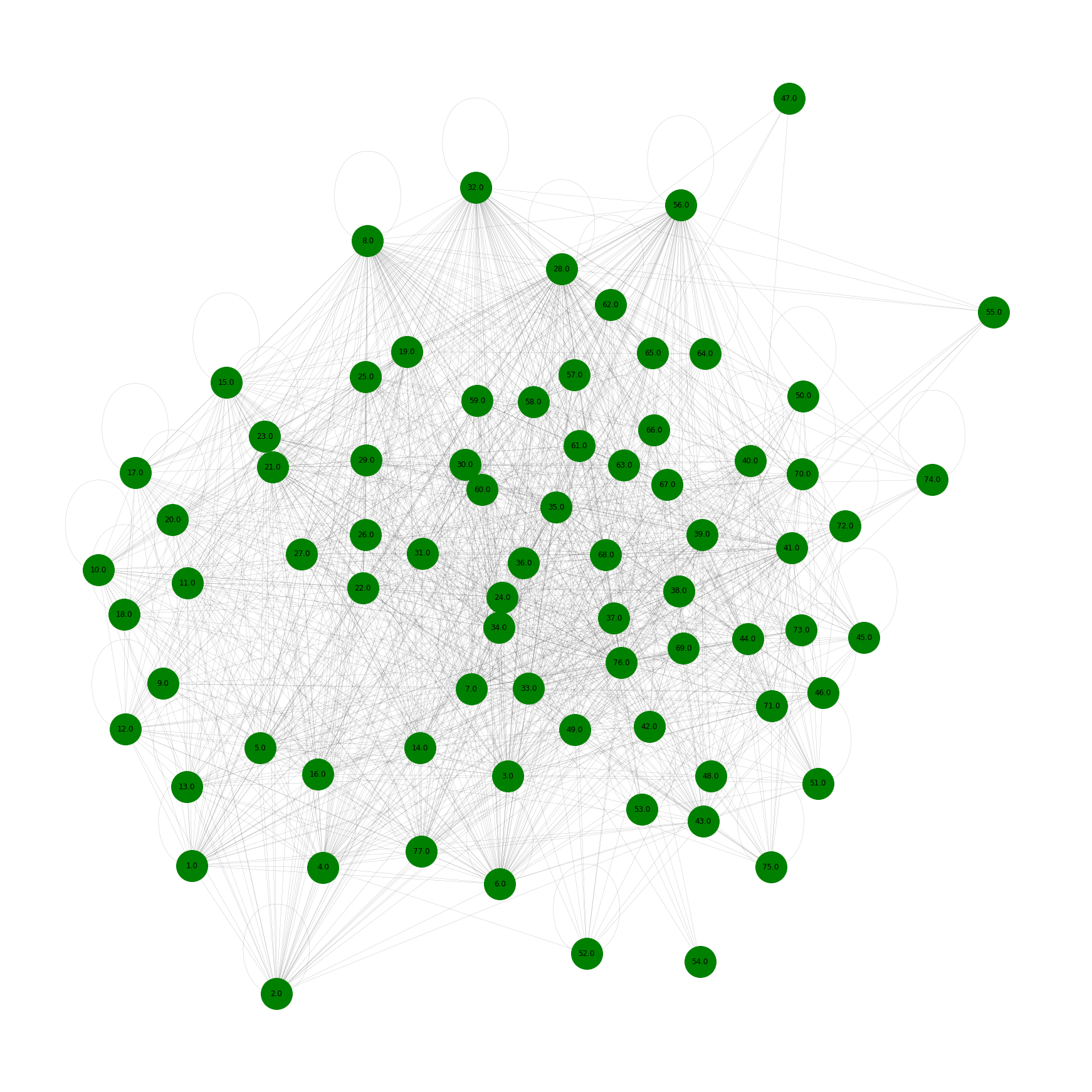}
    \caption{Unclustered Undirected Graph from taxi trip data}
    \label{fig:undir_first}
\end{subfigure} 
\hfill
\begin{subfigure}[b]{0.15\textwidth}
    \includegraphics[width=\textwidth]{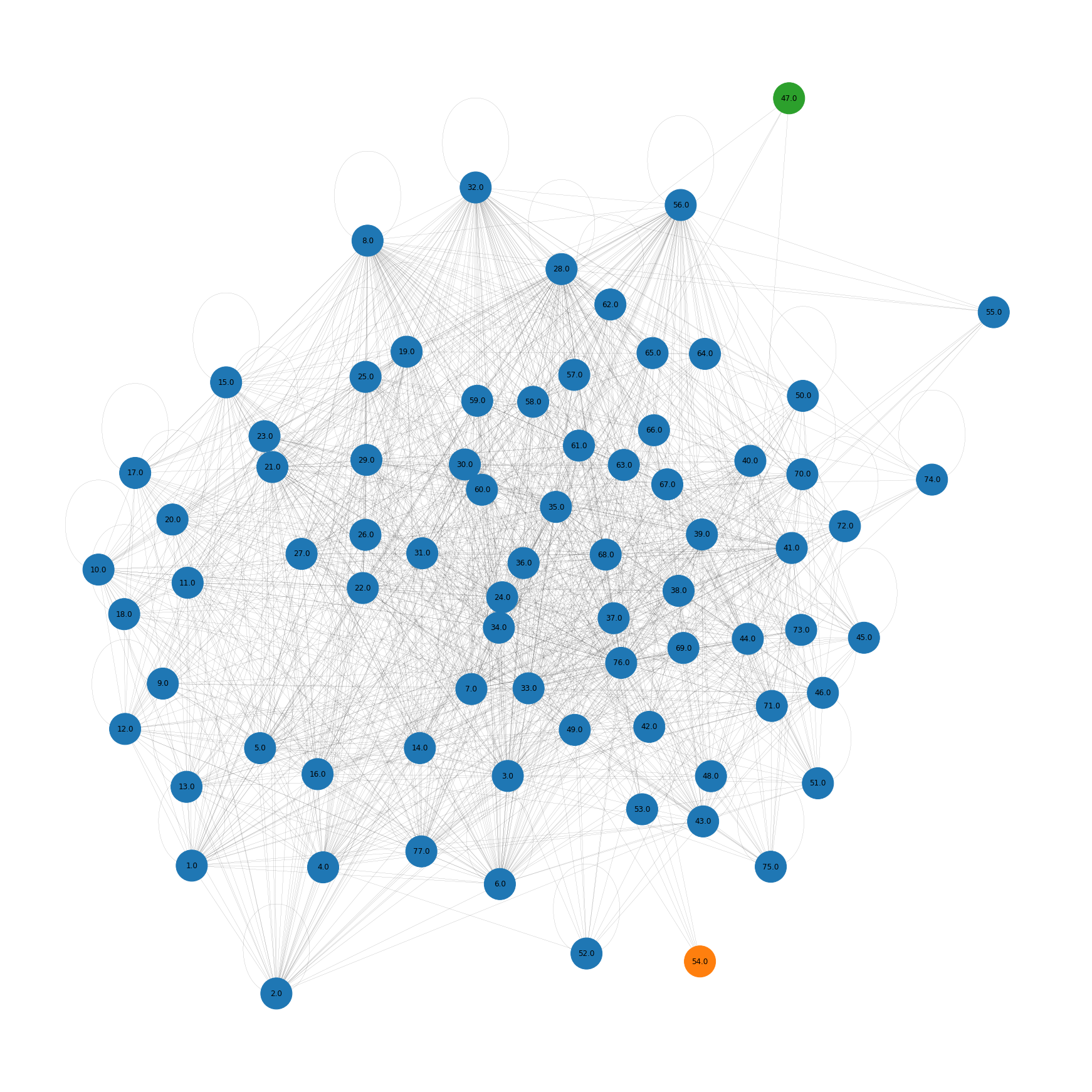}
    \caption{Spectral Clustering (unnormalized Laplacian)}
    \label{fig:undir_second}
\end{subfigure}
\hfill
\begin{subfigure}[b]{0.15\textwidth}
    \includegraphics[width=\textwidth]{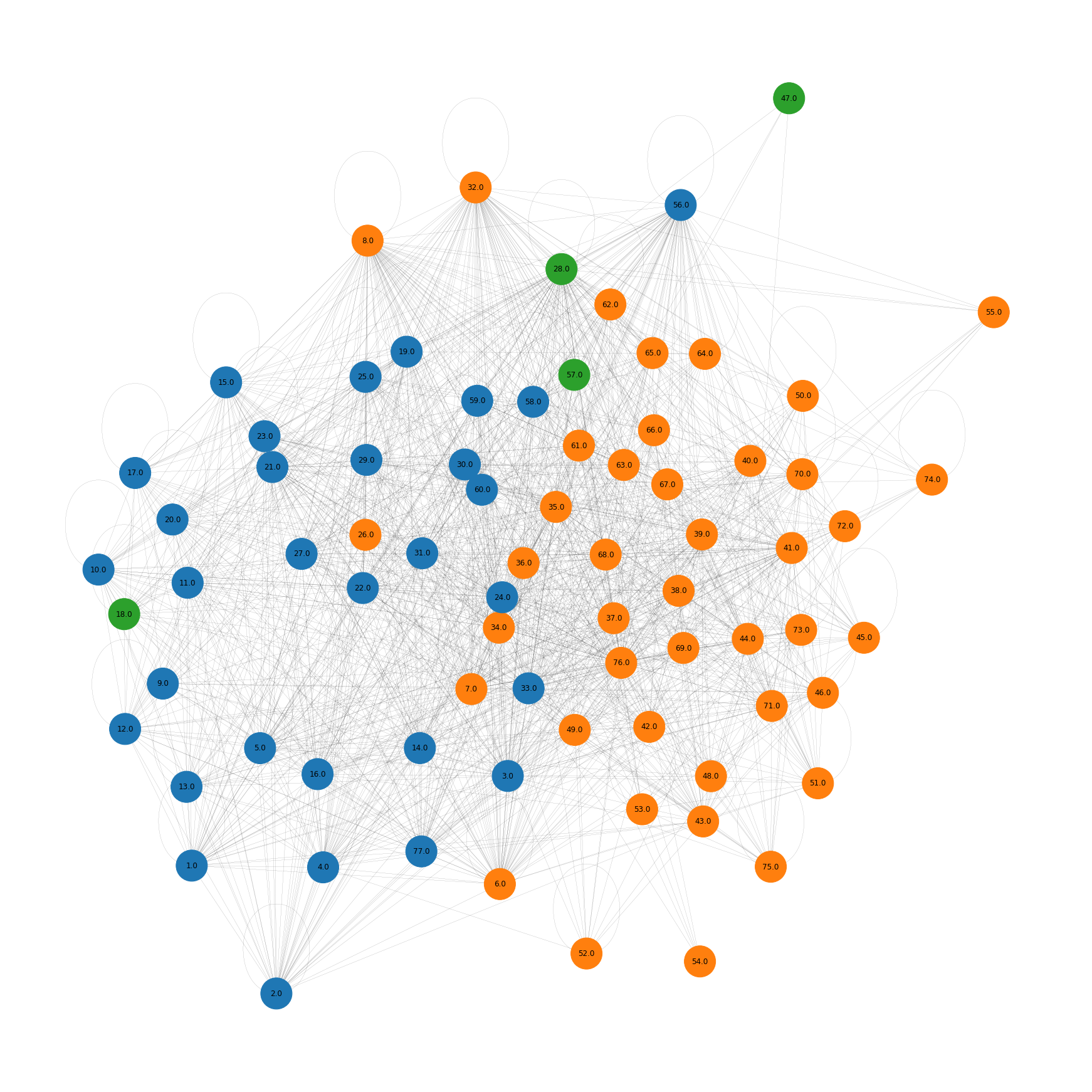}
    \caption{Spectral Clustering using normalized Laplacian}
    \label{fig:undir_third}
\end{subfigure}
\hfill
\begin{subfigure}[b]{0.15\textwidth}
    \includegraphics[width=\textwidth]{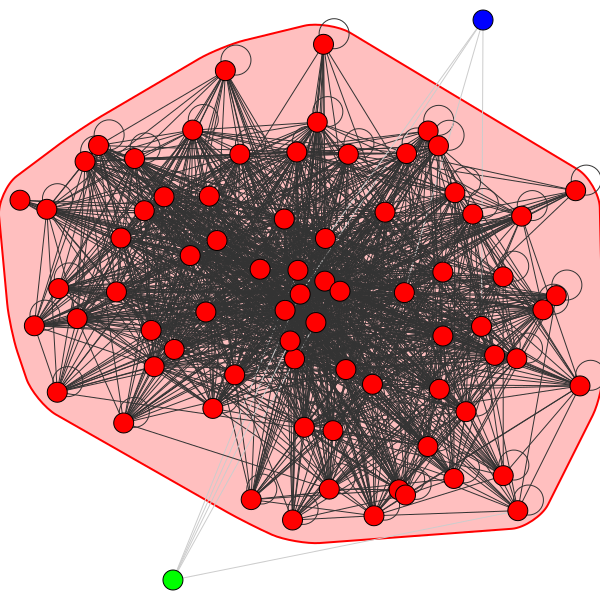}
    \caption{Clustered using Leiden algorithm}
    \label{fig:undir_fourth}
\end{subfigure}
\hfill
\begin{subfigure}[b]{0.15\textwidth}
    \includegraphics[width=\textwidth]{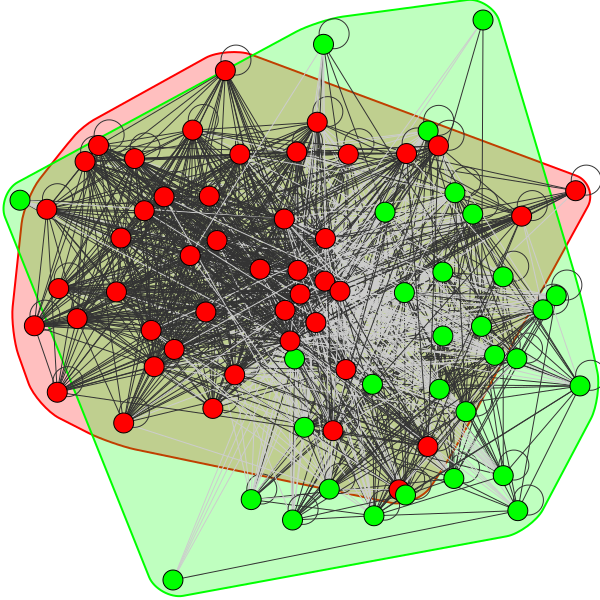}
    \caption{Clustered using Walktrap algorithm}
    \label{fig:undir_fifth}
\end{subfigure}

\caption{Spectral clustering algorithms for Undirected Graphs}
\label{fig:figure1}
\end{figure}

\begin{figure}[h]
\centering
\begin{subfigure}[b]{0.15\textwidth}
    \includegraphics[width=\textwidth]{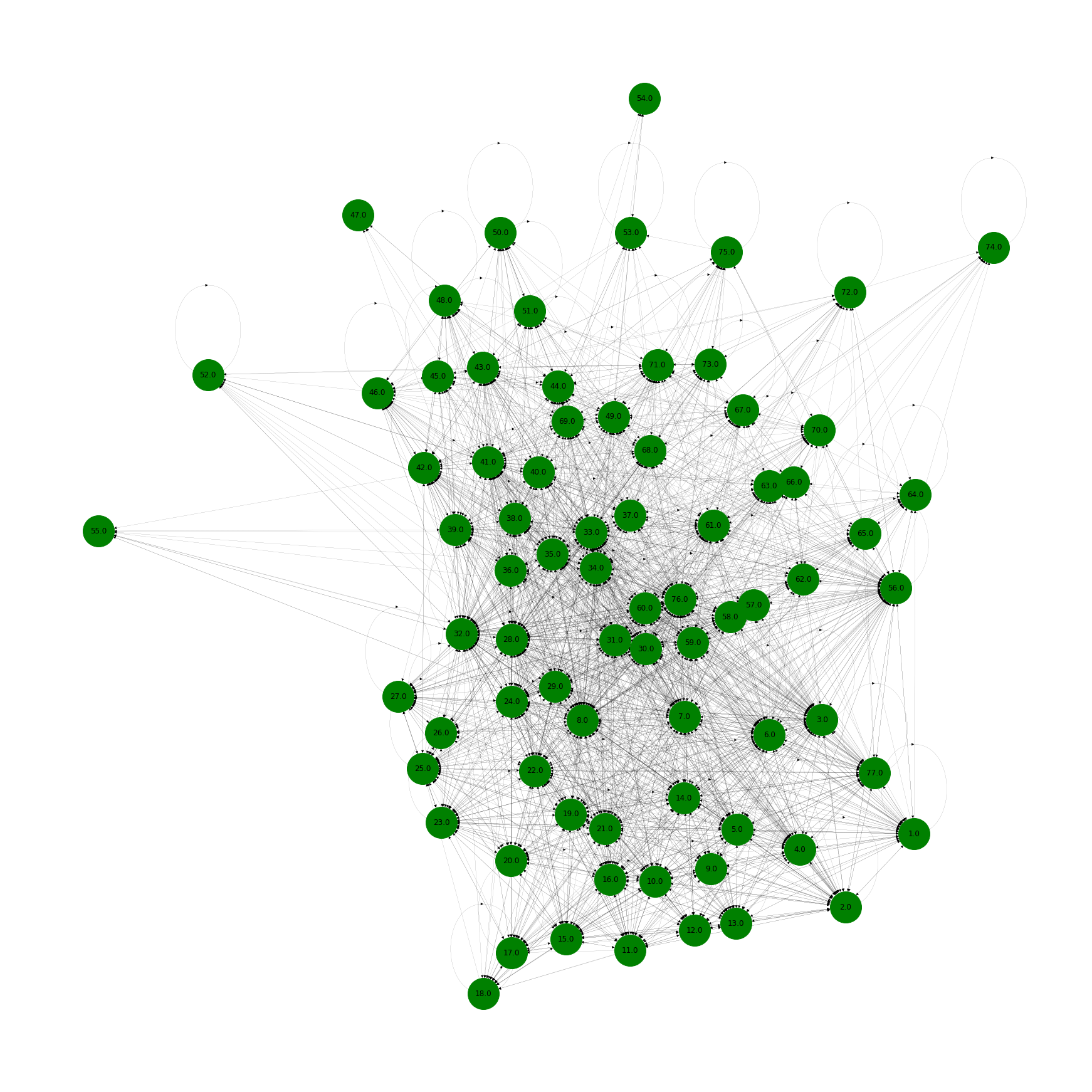}
    \caption{Unclustered Directed Graph from taxi trip data}
    \label{fig:dir_first}
\end{subfigure}
\hfill
\begin{subfigure}[b]{0.15\textwidth}
    \includegraphics[width=\textwidth]{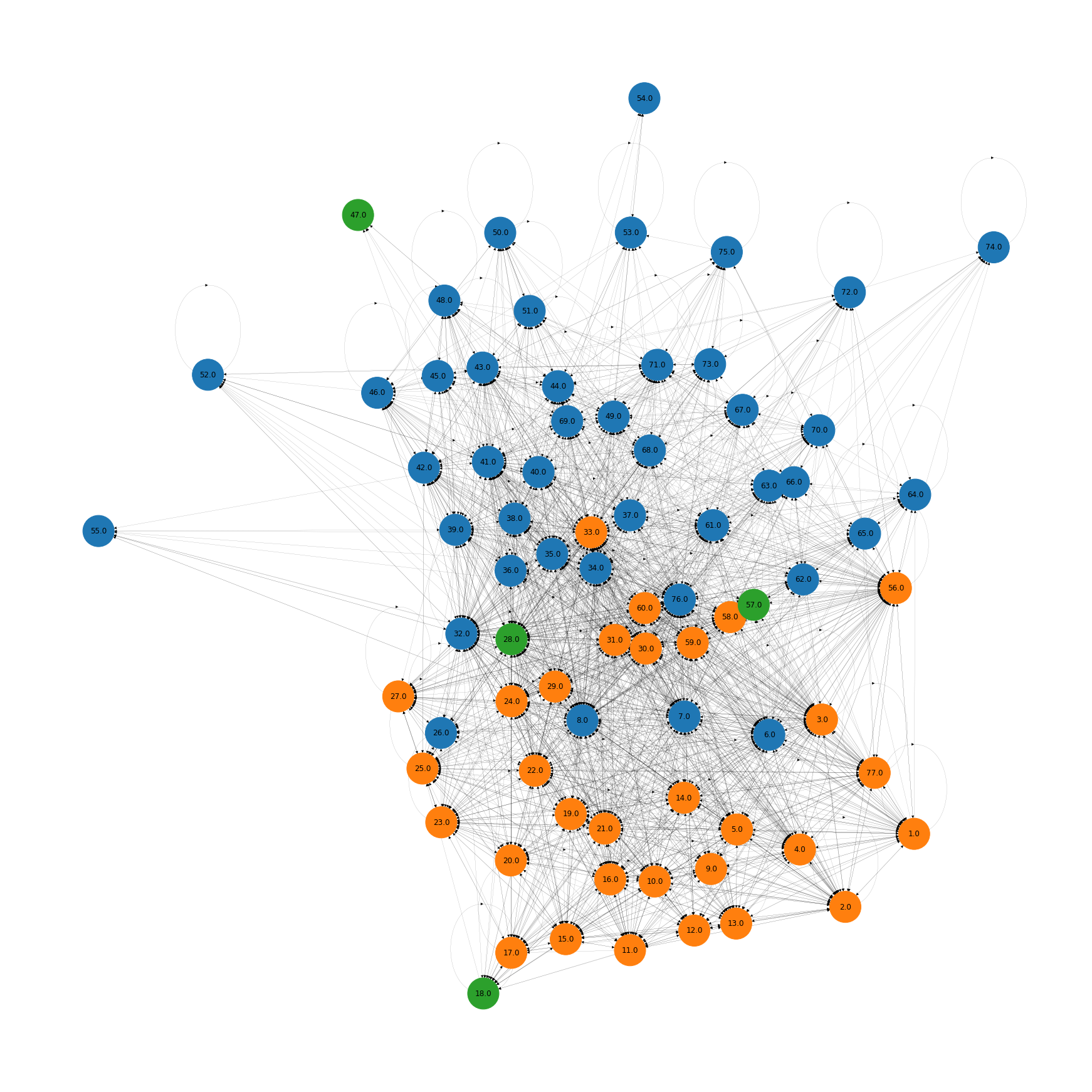}
    \caption{Clustered using Simple Transformation}
    \label{fig:dir_second}
\end{subfigure}
\hfill
\begin{subfigure}[b]{0.15\textwidth}
    \includegraphics[width=\textwidth]{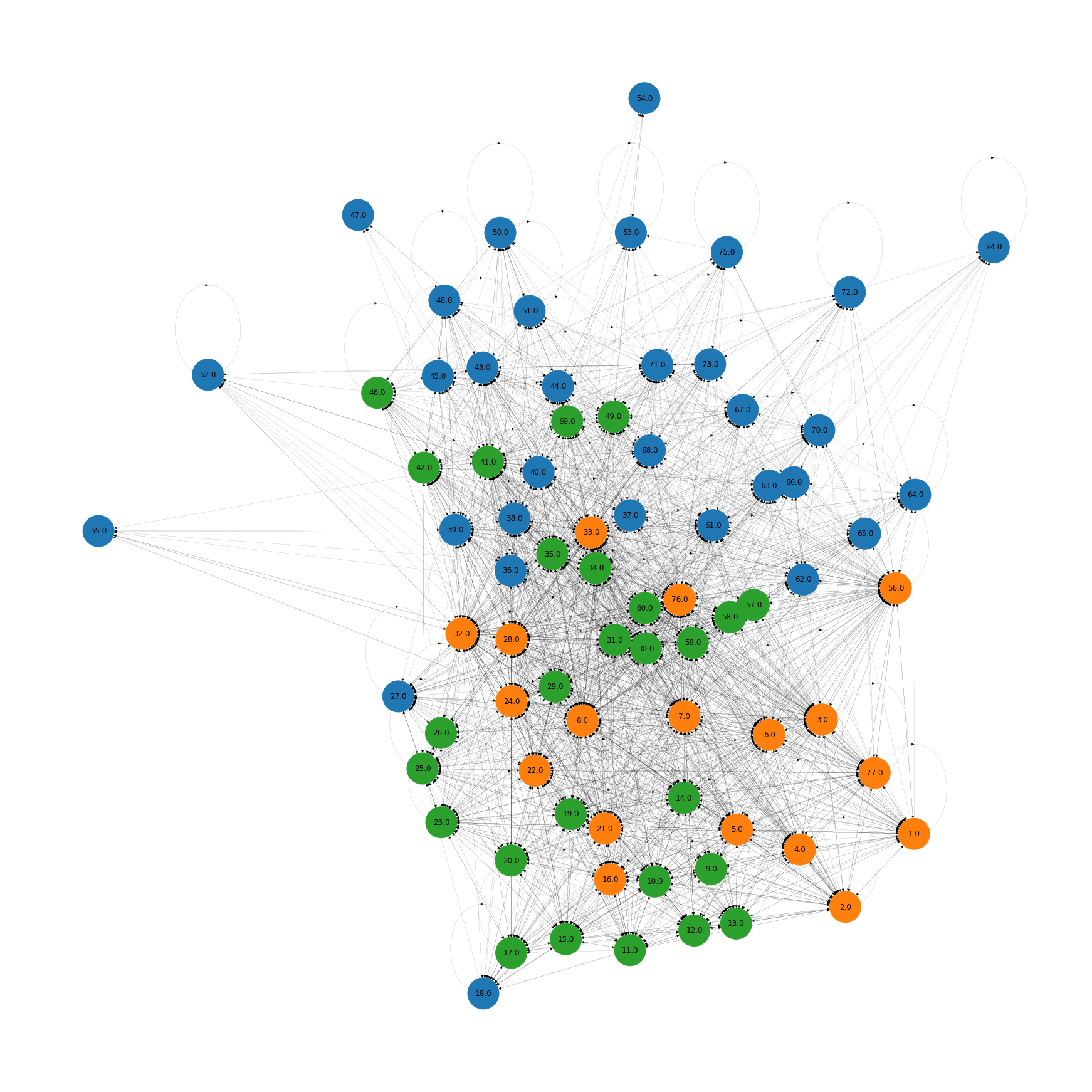}
    \caption{Clustered using Bibliometric Symmetrization}
    \label{fig:dir_third}
\end{subfigure}
\hfill
\begin{subfigure}[b]{0.15\textwidth}
    \includegraphics[width=\textwidth]{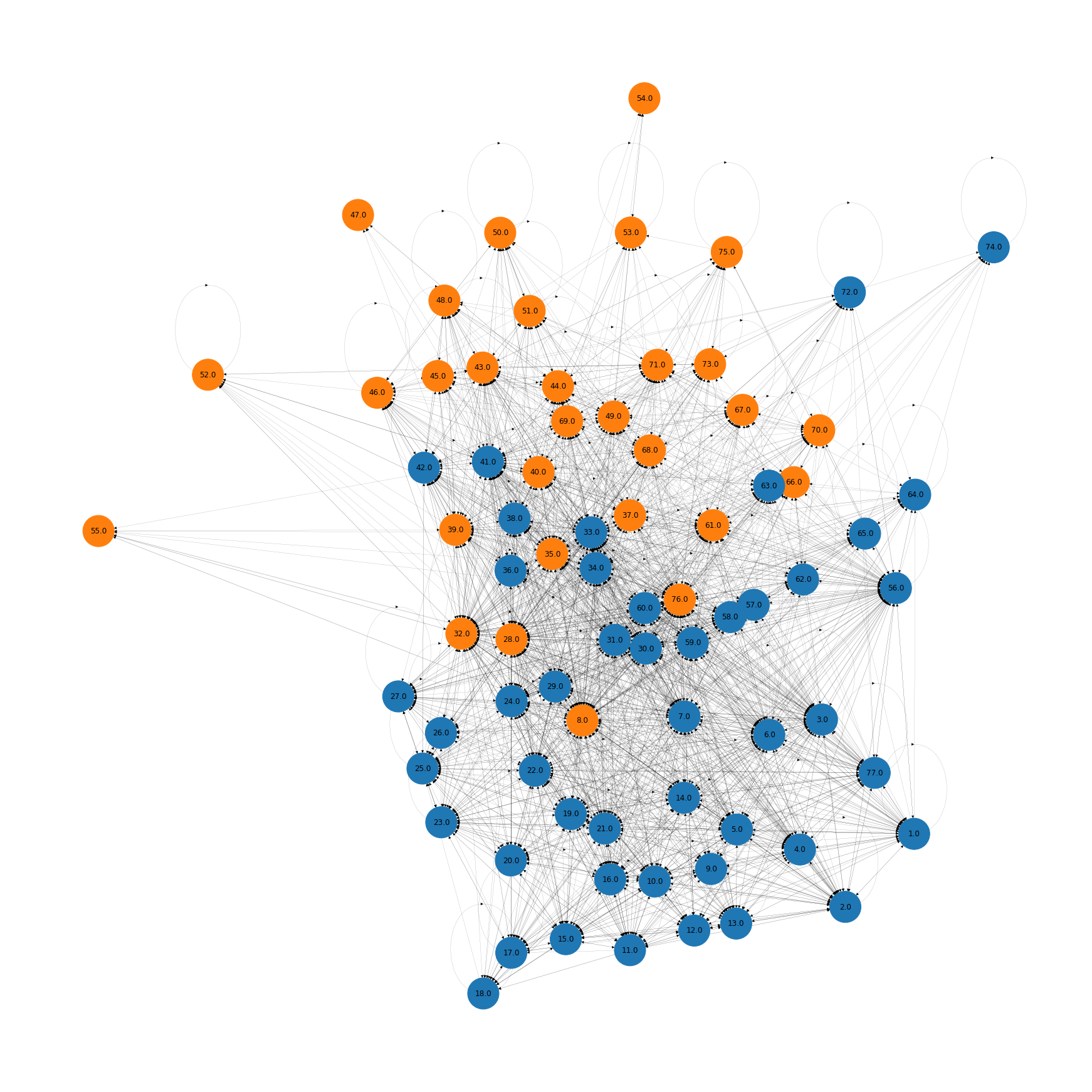}
    \caption{Clustered using Chung's Directed Laplacian}
    \label{fig:dir_fourth}
\end{subfigure}
\hfill
\begin{subfigure}[b]{0.15\textwidth}
    \includegraphics[width=\textwidth]{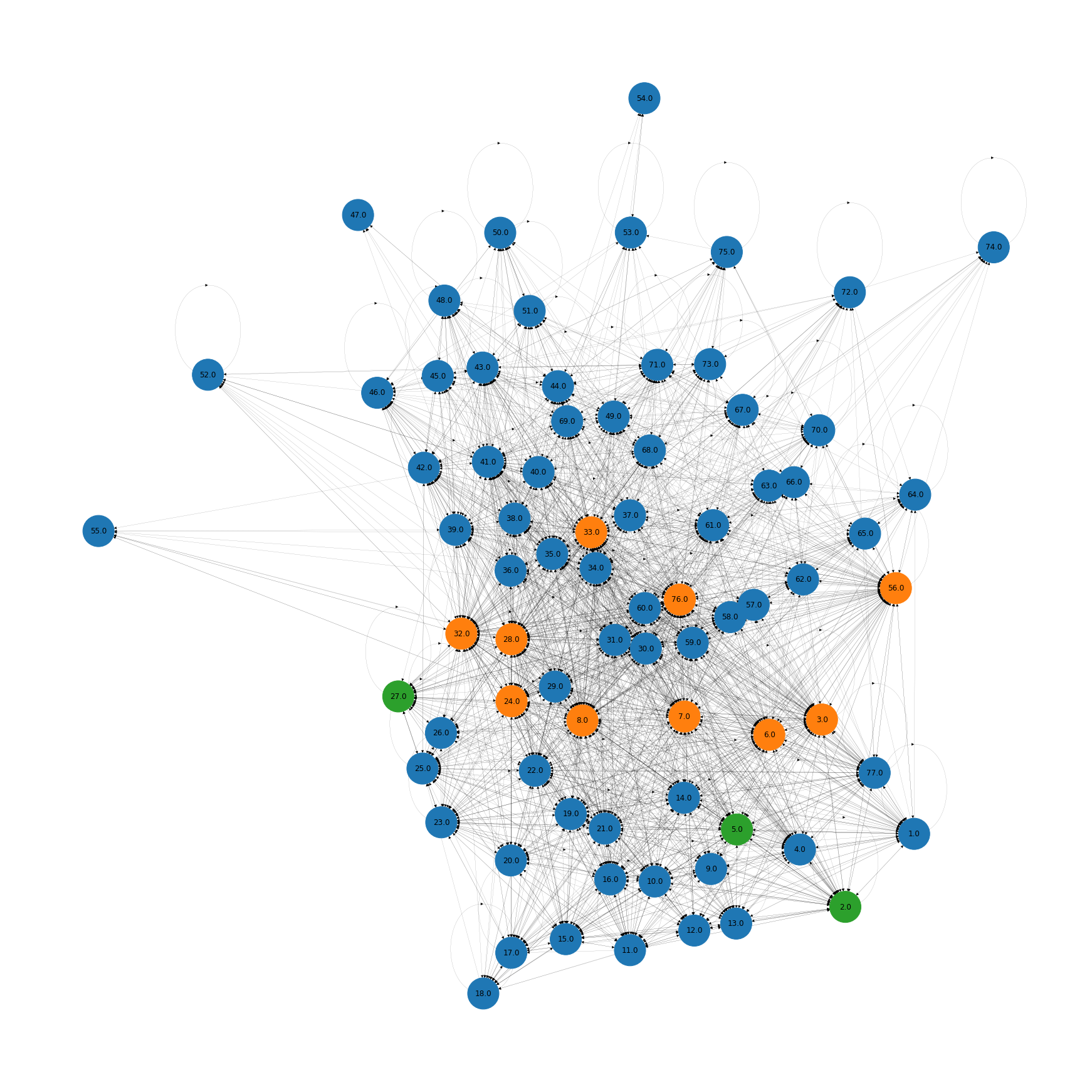}
    \caption{Clustered using SVD Spectral Clustering}
    \label{fig:dir_fifth}
\end{subfigure}
\hfill
\begin{subfigure}[b]{0.15\textwidth}
    \includegraphics[width=\textwidth]{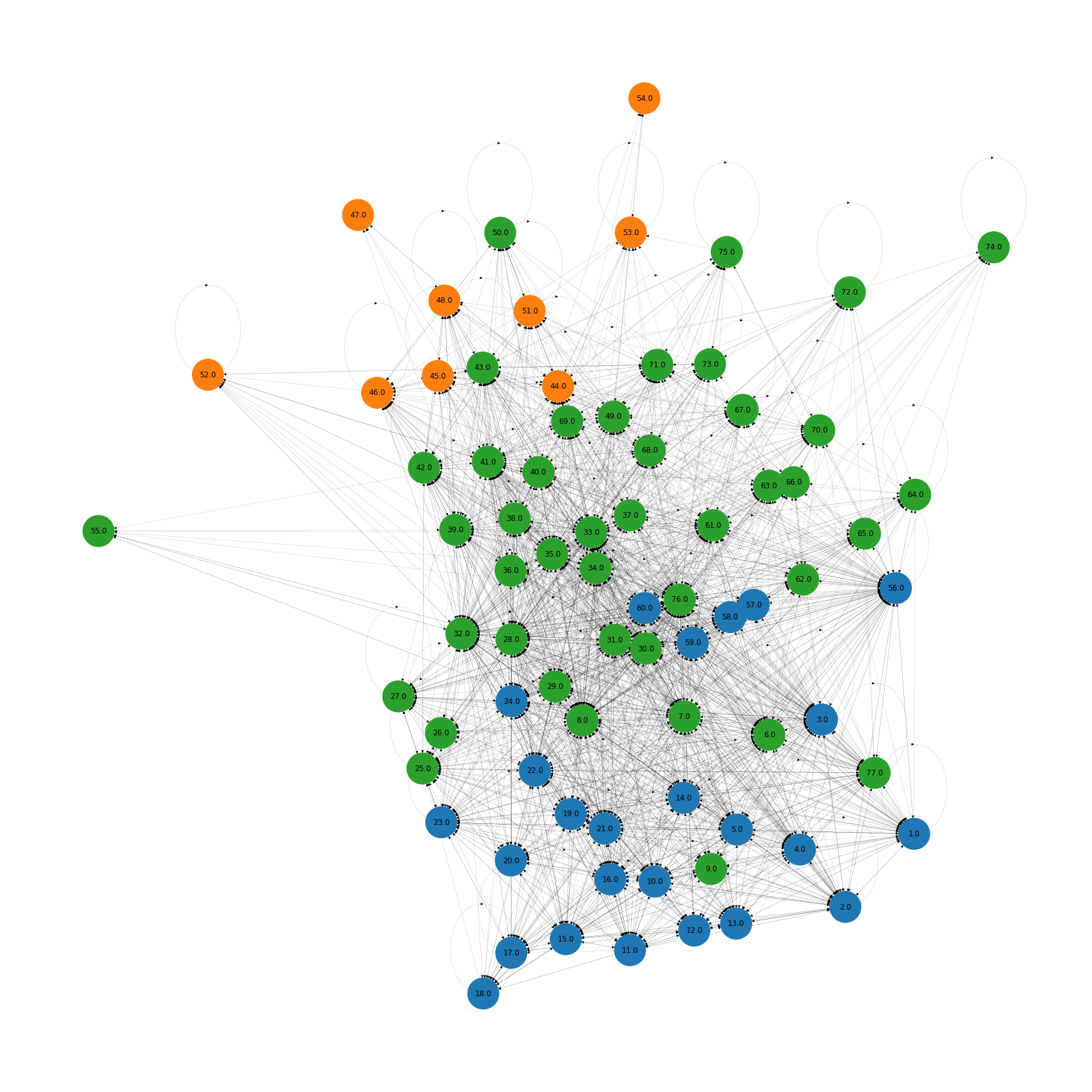}
    \caption{Spectral clustering algorithm based on random walk}
    \label{fig:dir_sixth}
\end{subfigure}

\caption{Spectral clustering algorithms for Directed Graphs}
\label{fig:figure2}
\end{figure}

The results of various clustering algorithms on undirected graphs are shown in Figure \ref{fig:figure1}. Neither spectral clustering with an unnormalized Laplacian nor clustering using the Leiden algorithm produced meaningful clusters, other than identifying outlier nodes. This is likely because the nodes were highly connected and the weights of the edges had a wide distribution. However spectral clustering with a normalized Laplacian and clustering using the Walktrap algorithm both found two clusters in the data. That two algorithms identified similar clusters indicates that there is some robustness to these results. For directed graphs (figure \ref{fig:figure2}), we can see that, for all the algorithms, one cluster can be identified that has nodes with a lot of incoming edges, whereas other clusters have relatively decreasing densities. All algorithms identified similar clusters indicating that there is some robustness to these results. 

In subsequent work, we could stratify the data by weekday/weekend in order to better understand the very different types of traffic flow that occur throughout the week. 

\newpage

\bibliography{aaai22}

\end{document}